\pageno=1                                      

\def\'#1{\ifx#1i{\accent"13\i}\else{\accent"13#1}\fi}

\def\v{{\bf v}}
\def\VS{V\'az\-quez-Se\-ma\-de\-ni}

\def\aa{Astron. Ap.\ }
\def\apj{{\refit  Astrophys.\ J.\/}\ }
\def\apjl{{\refit Astrophys.\ J.\ Lett.\/}\ }
\def\apjs{{\refit Astrophys.\ J.\ Suppl.\/}\ }
\def\apss{Ap. Sp. Sci.\ }
\def\mnras{MNRAS\ }

\def\ltsima{$\; \buildrel < \over \sim \;$}    
\def\lesssim{\lower.5ex\hbox{\ltsima}}           
\def\gtsima{$\; \buildrel > \over \sim \;$}    
\def\gtrsim{\lower.5ex\hbox{\gtsima}}           

\def\propsim{\lower.5ex\hbox{$\; \buildrel \propto \over \sim \;$}}

%
%
%
\font\ninerm=cmr9
\font\eightrm=cmr8
\font\sixrm=cmr6
\font\ninei=cmmi9
\font\eighti=cmmi8
\font\sixi=cmmi6
\skewchar\ninei='177 \skewchar\eighti='177 \skewchar\sixi='177
\font\ninesy=cmsy9
\font\eightsy=cmsy8
\font\sixsy=cmsy6
\skewchar\ninesy='60 \skewchar\eightsy='60 \skewchar\sixsy='60

\font\ninebf=cmbx9
\font\eightbf=cmbx8
\font\sixbf=cmbx6
\font\ninett=cmtt9
\font\eighttt=cmtt8
\hyphenchar\tentt=-1 
\hyphenchar\ninett=-1
\hyphenchar\eighttt=-1
\font\ninesl=cmsl9
\font\eightsl=cmsl8
\font\nineit=cmti9
\font\eightit=cmti8
\newskip\ttglue
\def\tenpoint{\def\rm{\fam0\tenrm}%
  \textfont0=\tenrm \scriptfont0=\sevenrm \scriptscriptfont0=\fiverm
  \textfont1=\teni \scriptfont1=\seveni \scriptscriptfont1=\fivei
  \textfont2=\tensy \scriptfont2=\sevensy \scriptscriptfont2=\fivesy
  \textfont3=\tenex \scriptfont3=\tenex \scriptscriptfont3=\tenex
  \def\it{\fam\itfam\tenit}%
  \textfont\itfam=\tenit
  \def\sl{\fam\slfam\tensl}%
  \textfont\slfam=\tensl
  \def\bf{\fam\bffam\tenbf}%
  \textfont\bffam=\tenbf \scriptfont\bffam=\sevenbf
   \scriptscriptfont\bffam=\fivebf
  \def\tt{\fam\ttfam\tentt}%
  \textfont\ttfam=\tentt
  \tt \ttglue=.5em plus.25em minus.15em
  \normalbaselineskip=12pt
  \let\sc=\eightrm
  \let\big=\tenbig
  \setbox\strutbox=\hbox{\vrule height8.5pt depth3.5pt width0pt}%
  \normalbaselines\rm}
\def\ninepoint{\def\rm{\fam0\ninerm}%
  \textfont0=\ninerm \scriptfont0=\sixrm \scriptscriptfont0=\fiverm
  \textfont1=\ninei \scriptfont1=\sixi \scriptscriptfont1=\fivei
  \textfont2=\ninesy \scriptfont2=\sixsy \scriptscriptfont2=\fivesy
  \textfont3=\tenex \scriptfont3=\tenex \scriptscriptfont3=\tenex
  \def\it{\fam\itfam\nineit}%
  \textfont\itfam=\nineit
  \def\sl{\fam\slfam\ninesl}%
  \textfont\slfam=\ninesl
  \def\bf{\fam\bffam\ninebf}%
  \textfont\bffam=\ninebf \scriptfont\bffam=\sixbf
   \scriptscriptfont\bffam=\fivebf
  \def\tt{\fam\ttfam\ninett}%
  \textfont\ttfam=\ninett
  \tt \ttglue=.5em plus.25em minus.15em
  \normalbaselineskip=10pt 
  \let\sc=\sevenrm
  \let\big=\ninebig
  \setbox\strutbox=\hbox{\vrule height8pt depth3pt width0pt}%
  \normalbaselines\rm}
\def\eightpoint{\def\rm{\fam0\eightrm}%
  \textfont0=\eightrm \scriptfont0=\sixrm \scriptscriptfont0=\fiverm
  \textfont1=\eighti \scriptfont1=\sixi \scriptscriptfont1=\fivei
  \textfont2=\eightsy \scriptfont2=\sixsy \scriptscriptfont2=\fivesy
  \textfont3=\tenex \scriptfont3=\tenex \scriptscriptfont3=\tenex
  \def\it{\fam\itfam\eightit}%
  \textfont\itfam=\eightit
  \def\sl{\fam\slfam\eightsl}%
  \textfont\slfam=\eightsl
  \def\bf{\fam\bffam\eightbf}%
  \textfont\bffam=\eightbf \scriptfont\bffam=\sixbf
   \scriptscriptfont\bffam=\fivebf
  \def\tt{\fam\ttfam\eighttt}%
  \textfont\ttfam=\eighttt
  \tt \ttglue=.5em plus.25em minus.15em
  \normalbaselineskip=9pt
  \let\sc=\sixrm
  \let\big=\eightbig
  \setbox\strutbox=\hbox{\vrule height7pt depth2pt width0pt}%
  \normalbaselines\rm}
%
\def\headtype{\ninepoint}                 
\def\abstracttype{\ninepoint}             
\def\captiontype{\ninepoint}              
\def\footnotetype{\ninepoint}             
\def\refit{\it}                           
\font\chaptitle=cmr10 at 11pt             
\rm                                       

%
%
\parindent=0.25in                         
\parskip=0pt                              
\baselineskip=12pt                        
\hsize=4.25truein                         
\vsize=7.445truein                        
\hoffset=1in                              
\voffset=-0.5in                           

\newskip\sectionskipamount                
\newskip\aftermainskipamount              
\newskip\subsecskipamount                 
\newskip\firstpageskipamount              
\newskip\capskipamount                    
\newskip\ackskipamount                    
\sectionskipamount=0.2in plus 0.09in
\aftermainskipamount=6pt plus 6pt         
\subsecskipamount=0.1in plus 0.04in
\firstpageskipamount=3pc
\capskipamount=0.1in
\ackskipamount=0.15in
\def\sectionskip{\vskip\sectionskipamount}
\def\aftermainskip{\vskip\aftermainskipamount}
\def\subsecskip{\vskip\subsecskipamount} 
\def\firstpageskip{\vskip\firstpageskipamount}
\def\capskip{\hskip\capskipamount}

%
%
\nopagenumbers                            
\newcount\firstpageno                     
\firstpageno=\pageno                      
\newcount\chapno                          

\def\rightheadline{\headtype\phantom{\folio}\hfil\runningtitletext\hfil\folio}
\def\leftheadline{\headtype\folio\hfil\runningauthortext\hfil\phantom{\folio}}
\headline={\ifnum\pageno=\firstpageno\hfil
           \else
              \ifdim\ht\topins=\vsize           
                 \ifdim\dp\topins=1sp \hfil     
                 \else
                     \ifodd\pageno\rightheadline\else\leftheadline\fi
                 \fi
              \else
                 \ifodd\pageno\rightheadline\else\leftheadline\fi
              \fi
           \fi}

\def\bottomnumber{\hss\tenrm[\folio]\hss}
\footline={\ifnum\pageno=\firstpageno\bottomnumber\else\hfil\fi}

%
%
%
%
\outer\def\mainsection#1
    {\vskip 0pt plus\smallskipamount\sectionskip
     \message{#1}\vbox{\noindent{\bf#1}}\nobreak\aftermainskip\noindent}
 
\outer\def\subsection#1
    {\vskip 0pt plus\smallskipamount\subsecskip
     \message{#1}\vbox{\noindent{\bf#1}}\nobreak\smallskip\nobreak\noindent}
 
\def\backup{\nobreak\vskip-\baselineskip\nobreak\vskip-\subsecskipamount\nobreak
}

\def\title#1{{\chaptitle\leftline{#1}}}
\def\name#1{\leftline{#1}}
\def\affiliation#1{\leftline{\it #1}}
\def\abstract#1{{\abstracttype \noindent #1 \smallskip\vskip .1in}}
\def\ref{\noindent \parshape2 0truein 4.25truein 0.25truein 4truein}
\def\caption{\noindent \captiontype
             \parshape=2 0truein 4.25truein .125truein 4.125truein}

\def\footnote#1{\edef\fspafac{\spacefactor\the\spacefactor}#1\fspafac
      \insert\footins\bgroup\footnotetype
      \interlinepenalty100 \let\par=\endgraf
        \leftskip=0pt \rightskip=0pt
        \splittopskip=10pt plus 1pt minus 1pt \floatingpenalty=20000
        \textindent{#1}\bgroup\strut\aftergroup\strut\egroup\let\next}
\skip\footins=12pt plus 2pt minus 4pt 
\dimen\footins=30pc 

%
%

\def\@{\spacefactor 1000}

\def\,{\pcomma} 
\def\pcomma{\relax\ifmmode\mskip\thinmuskip\else\thinspace\fi}

\def\oversim#1#2{\lower0.5ex\vbox{\baselineskip=0pt\lineskip=0.2ex
     \ialign{$\mathsurround=0pt #1\hfil##\hfil$\crcr#2\crcr\sim\crcr}}}

\def\runningtitletext{MHD Turbulence}
\def\runningauthortext{V\'azquez-Semadeni et al.}

\null
\firstpageskip

{\baselineskip=14pt
\title{COMPRESSIBLE MHD TURBULENCE: IMPLICATIONS FOR}
\title{MOLECULAR CLOUD AND STAR FORMATION
}

\vskip .3truein
\name{ENRIQUE VAZQUEZ-SEMADENI}
\affiliation{Instituto de Astronom\'ia, UNAM}
\vskip .1truein
\name{EVE C. OSTRIKER}
\affiliation{Astronomy Dept., University of Maryland}
\vskip .1truein
\name{THIERRY PASSOT}
\affiliation{Observatoire de Nice}
\vskip .1truein
\name{CHARLES F. GAMMIE}
\affiliation{Harvard-Smithsonian CFA}
\leftline{and}
\vskip .1truein
\name{JAMES M. STONE}
\affiliation{Astronomy Dept., University of Maryland}
\vskip .3truein

\abstract{We review recent results from numerical simulations and
related models of MHD turbulence in the interstellar medium (ISM) and
in molecular clouds. We discuss the implications of turbulence for the
processes of cloud formation and evolution, and the determination of
clouds' physical properties. Numerical simulations of the turbulent
ISM to date have included magnetic fields, self-gravity, parameterized
heating and cooling, modeled star formation and other turbulent
inputs.  The structures which form reproduce well observed
velocity-size scaling properties, while predicting the non-existence
of a general density-size scaling law.  Criteria for the formation of
gravitationally-bound structures by turbulent compression are
summarized. For flows with equations of state $P\propto
\rho^\gamma$, the statistics of the density field depend on the
exponent $\gamma$.  Numerical simulations of both forced and decaying
MHD compressible turbulence have shown that the decay rate is
comparable to the non-magnetic case.  For virialized clouds, the
turbulent decay time is shorter than the gravitational free-fall time,
so wholesale cloud collapse is only prevented by ongoing turbulent
inputs and/or a strong mean magnetic field. Finally, perspectives for
future work in this field are briefly discussed.}

\vfil\eject

\mainsection{I.~~INTRODUCTION}
\backup
\subsection{A.~~Observational Motivation}

Present-day star formation in our Galaxy is observed to take place in
cold molecular clouds, which appear to be in a state of highly
compressible MHD turbulence.  Furthermore, the atomic gas, within
which the molecular clouds form, is also turbulent on larger scales,
from hundreds to thousands of pc (e.g., Braun 1998).  In this chapter
we review recent results in the areas of cloud formation, structure
and evolution, as well as their implications for observed physical and
statistical cloud properties; these results are obtained mainly from
numerical simulations of compressible MHD turbulence, and related
analytical models. An introductory review including a more detailed
discussion of turbulence basics and discussions (current up to 1997)
on fractality and phenomenological models has recently been given by
V\'azquez-Semadeni (1999).  The review of Ostriker (1997) focuses on
the role of MHD turbulence in the internal dynamical evolution of
molecular clouds.  The reviews of Heiles et al. (1993) and McKee et al.
(1993) provide extensive discussions of the observations and general
theory, respectively, of magnetic fields in star-forming regions.

The formation of molecular clouds probably cannot be considered
separately from the formation of larger diffuse HI structures, since
the former are often observed to have HI ``envelopes'' (e.g., Peters
\& Bash 1987; Shaya \& Federman 1987; Wannier et al.\ 1991; Blitz
1993; Williams et al.\ 1995), suggesting that molecular clouds and
clumps can be regarded as the ``tips of the icebergs'' in the general
continuum interstellar density field of galaxies.  The process of
cloud formation quite possibly involves more than a single mechanism,
including the passage of spiral density waves and the effects of
combined large-scale instabilities (e.g., Elmegreen 1993a, 1995;
Gammie 1996) operating preferentially in the formation of the largest
high-density structures, and the production of smaller density
condensations by either swept-up shells (Elmegreen \& Elmegreen 1978;
Vishniac 1983, 1994; Hunter et al.\ 1986), or by a generally turbulent
medium (Hunter 1979; Hunter \& Fleck 1982; Tohline et al.\ 1987;
Elmegreen 1993b). In Sec.\ II of this chapter we discuss recent
results on the generation of density fluctuations in a turbulent
medium, obtained from numerical simulations and semi-phenomenological
models. Other mechanisms of cloud formation have been reviewed
extensively by Elmegreen (1993a, 1995).  Note, however, that discrete
cloud coagulation mechanisms discussed there may not be directly
applicable in the context of a dynamically-evolving continuum, such as
that considered here.

Structurally, molecular clouds are very complex, with volume- averaged
H$_2$ densities $n($H$_2)$ ranging from $\lesssim 50$ cm$^{-3}$ for
giant molecular clouds of sizes several tens of pc (e.g., Blitz 1993),
to $n($H$_2) \gtrsim 10^5$ cm$^{-3}$ for dense cores of sizes
0.03--0.1 pc (e.g., Wilson \& Walmsley 1989; Plume et al.\ 1997;
Pratap et al.\ 1997; Lada et al.\ 1997; Myers 1995).  Kinetic
temperatures are relatively constant, typically $T \sim 10$ K $\pm$ a
few degrees (e.g., Pratap et al.\ 1997), although temperatures larger
by factors of a few are found in the vicinity of star-forming regions
(e.g., Torrelles et al.\ 1983; Solomon \& Sanders 1985). Their
internal velocity dispersions are generally supersonic, with Mach
numbers up to $\gtrsim 10$, except for the smallest cores (size $R
\lesssim 0.1$ pc) (e.g., Larson 1981; Blitz 1993 and references
therein; Goodman et al.1998). When available (and significant; see,
e.g., Crutcher et al. 1993, 1996; Padoan \& Nordlund 1998a), Zeeman
measurements give typical values of the magnetic field intensity of a
few to a few tens of $\mu$G (e.g., Heiles et al.\ 1993; Crutcher et al.
1993; Troland et al. 1996; Crutcher 1998), consistent with near
equipartition between the kinetic and magnetic energies.

Additionally, interstellar clouds seem to follow power-law scaling
laws between their average density, velocity dispersion and size
(Larson 1981; see also Blitz 1993 and references therein). Together
with the near equipartition between kinetic and magnetic energy, these
are normally interpreted as evidence for virialized magnetic support
for the clouds (e.g., Shu et al.\ 1987; Mouschovias 1987; Myers \&
Goodman 1988a,b; see also Whitworth 1996), although they only provide
arguments of self-consistency of the virial equilibrium hypothesis
rather than conclusive proofs (Heiles et al.\ 1993). In addition to
the clouds that are self-consistent with virialization, however,
examples also exist of objects which appear to be highly disturbed
(e.g., Carr 1987; Loren 1989; Plume et al.\ 1997), or simply regions
away from map intensity maxima (Falgarone et al.\ 1992, 1998), which
do not satisfy one or more of the scalings. Another example of such
scalings is the mass spectrum of the clouds and clumps, which seems to
also be a power-law (e.g., Blitz 1993; Williams et al. 1995), although
present Galactic cloud identification surveys remain incomplete.

Molecular clouds also exhibit signatures of self-similarity, or more
generally, multifractality. Scalo (1990) and Falgarone et al.\ (1991)
have shown that cloud boundaries have projected fractal dimensions
$\sim 1.4$. Also, clouds exhibit hierarchical structure; i.e., the
densest structures are nested within larger, less dense ones in a
self-similar fashion (e.g., Scalo 1985), at least at large scales. The
self-similarity appears to either break down or at least change
similarity exponents at some small scale, reported by Larson (1995) to
be close to $0.05$ pc in Taurus, corresponding to the local Jeans
length, although the latter point is controversial (e.g., Simon 1997;
Bate et al.\ 1998). A similar break has also been reported at scales
0.25--0.5 pc (Blitz \& Williams, 1997), showing that this question is
still largely open.

Such complexity strongly suggests itself as a manifestation of the
turbulent regime which permeates molecular clouds. Thus, an
understanding of the basics of compressible MHD turbulence and its
incarnation in the interstellar case is necessary for explaining
molecular cloud structure and evolution, and for diagnosing the
consequences of turbulent dynamics for the process of star formation.
The disordered, nonlinear nature of interstellar MHD turbulence makes
direct numerical simulation the most useful tool for developing this
understanding.  In Sec.\ III of this chapter, we summarize recent
results on cloud structure and discuss questions such as whether the
physical conditions and scaling laws observed in clouds arise
naturally in simulations of the turbulent ISM, and what simulations
predict for energy balance in the clouds, density fluctuation
statistics, and the general characterization of structure in these
flows.

Evolutionary aspects of turbulence and the mechanisms of cloud support
under highly nonlinear, strongly self-gravitating conditions have
remained open questions since the original identification of turbulent
molecular clouds in the ISM.  Supersonic turbulent maintenance faces
the well-known problem of an expected excessive dissipation in shocks
(Goldreich \& Kwan 1974). Arons and Max (1975) proposed that the
``turbulent'' motions in molecular clouds may actually be
moderate-amplitude (sub-Alfv\'enic) MHD waves, arguing that shock
formation and the associated dissipation would then be significantly
diminished.  Additionally, it has been suggested that even strong
dissipation may be compensated by sufficient energy injection from
embedded stars and other sources (e.g. Norman and Silk 1980; Scalo
1987; McKee 1989; Kornreich \& Scalo 1998).  Sec.\ IV in this chapter
discusses recent numerical MHD results on calculating dissipation
rates for turbulence with parameters appropriate for the cold ISM, and
the implications of these results for questions of cloud support
against self-gravitating contraction.  In Sec.\ V we conclude with a
summary and discussion of outstanding goals and challenges for future
research.

\subsection{B.~~MHD Compressible Turbulence Basics}

Before proceeding to the next sections, a few words on the nature and
parameters of molecular cloud turbulence are in order. Its properties
are very different from laboratory incompressible turbulence and in
fact additional non-dimensional parameters are necessary to
characterize the flow besides the viscous and magnetic Reynolds
numbers (which respectively measure the magnitude of the nonlinear
advection term $\v \cdot \nabla \v$ in the momentum equation with
respect to the viscous and Ohmic dissipation terms). Two of these
parameters are the sonic (thermal) and Alfv\'enic Mach numbers
$M_s\equiv u/c$ and $M_A\equiv u/v_A$, where $u$ is a characteristic
velocity of the flow, $c$ is the isothermal sound speed and $v_A=B/(4
\pi \rho)^{1/2}$ is the Alfv\'en speed, with $B$ the magnetic field
strength and $\rho$ the fluid density. The plasma beta, $\beta\equiv
c^2\rho/(B^2/8 \pi)=2(M_A/M_s)^2$ (i.e., the ratio of the thermal to
the magnetic pressure) is also frequently used to characterize the
importance of magnetic fields.  The last basic parameter, which
characterizes the importance of self-gravity, is the Jeans number
$n_J\equiv L/L_J=L/\sqrt{\pi c^2/G \rho}$ (i.e. this compares the
cloud linear scale $L$ with the minimum Jeans-unstable wavelength
$L_J$).

Other peculiarities of ISM turbulence are related to the driving
mechanisms. In standard (``Kolmogorov'') incompressible turbulence
theory, energy is 
assumed to be injected at large scales; from there it cascades down to
the small scales where it is dissipated.  In the ISM, forcing occurs
at all scales (e.g., Scalo 1987; Norman \& Ferrara 1996): some
mechanisms operate at large scales ($\gtrsim 1$ kpc), such as the
Galactic shear (Fleck 1981, although this is believed to be a very
inefficient source; see, e.g., Shu et al. 1987) and supershells;
others at intermediate scales ($\gtrsim 100$ pc), such as expanding
HII regions and supernova explosions. Yet others act at small scales
(a few tenths to a few pc), such as stellar winds or bipolar
outflows. All these mechanisms are important sources of kinetic
energy.

Small-scale dissipation mechanisms in the cold ISM are of several
types.  First, there is the usual viscous dissipation operating mostly
in shocks and which remains finite even in the limit of vanishing
viscosity.  A significant amount of dissipation also results from
ambipolar diffusion due to ion-neutral friction at high densities in
the presence of a strong enough magnetic field (Kulsrud and Pearce
1969; see McKee et al. 1993; Myers and Khersonsky 1995).  Cooling
processes are also very efficient and may radiate much of the
dissipated energy, but are balanced by heating processes and may
result in near-isothermal or quasi-polytropic behavior (i.e. $P\propto
\rho^\gamma$) (Sec.\ II).

Even aside from all these astrophysical properties, compressible
turbulence has many distinctive features related to the transfer of
energy from large scales to the dissipation (thermalization)
scale. The new ingredient with respect to incompressible turbulence is
the existence of {\it potential} (or compressible) modes, in addition
to the solenoidal (vortical) ones.  Because compressions and
rarefactions can exchange energy between bulk kinetic and microscopic
thermal parts, compressible flows do not in general conserve bulk
kinetic energy.  Strongly radiative shocks, which yield an
irreversible energy loss, are the most important aspect of this
feature.  The presence of additional compressive and thermal degrees
of freedom therefore excludes the use of dimensional analysis to
determine the slope of the velocity spectrum, as is done (by
assumption of an energy-conservative cascade through scales) in the
so-called K41 theory (Kolmogorov 1941; Obukhov 1941; see e.g. Landau
and Lifshitz 1987), which produces the well-known ``universal''
spectrum of the form $du^2/dk\equiv E(k) \propto k^{-5/3}$ for
incompressible, non-magnetic turbulence, where $k=2 \pi/\lambda$ is
the wavenumber associated to wavelength $\lambda$.  For magnetized,
incompressible, strong turbulence, recent theoretical work suggests
that the mean magnetic field leads to strong anisotropy in the cascade
but the same averaged energy spectrum (Goldreich and Sridhar 1995).

Moreover, the K41 theory is based on the assumption of locality in
Fourier space of the nonlinear cascade (i.e., transfer between Fourier
modes of similar wavelengths), a hypothesis probably not valid in the
compressible case because coupling among very different scales occurs
in shocks.  In these highly intermittent (i.e., inhomogeneous in space
and time) structures, all Fourier modes decay at the same rate
(Kadomtsev \& Petviashvili 1973; Landau \& Lifshitz 1987), thus
invalidating the notion of inviscid cascade along the ``inertial
range'', defined as the range in Fourier space where the energy flux
is constant.  Note however that cascade processes in the presence of
shocks have been discussed by Kraichnan (1968) and Kornreich \& Scalo
(1998).  The spectrum of the compressible modes appears to approach a
Burgers (1974) spectrum of the form $k^{-2}$, which arises simply from
the Fourier structure of the shocks. The corresponding
configuration-space scaling for Burgers turbulence is $u_l\propto
l^{1/2}$ (sec.\ III.A).  This has been observed in numerical
simulations of strongly compressible turbulence even in the presence
of the magnetic field (Passot, Pouquet \& Woodward 1988; Passot, \VS\
\& Pouquet 1995; Gammie \& Ostriker 1996; Balsara, Crutcher, \& Pouquet 
1997; Stone 1998).

\mainsection{{I}{I}.~~TURBULENT CLOUD FORMATION}

An important question is whether structures formed by either turbulent
compressions or passages of single shock waves can become
gravitationally unstable and collapse (\"Ogelman \& Maran 1976;
Elmegreen \& Lada 1977; Elmegreen \& Elmegreen 1978; Hunter 1979;
Hunter \& Fleck 1982; Hunter et al.\ 1986; Tohline et al.\ 1987;
Stevens et al.\ 1992; Elmegreen 1993b).  As pointed out by Hunter et
al.\ (1986), when both heating and cooling are present, the isothermal
approximation, often used to describe radiative flows, is just one out 
of a continuum of possibilities. In cases where the heating and
cooling rates are 
reasonably approximated by power-law functions of the density and
temperature, and faster than the dynamical rates, the gas can be
described as a barotropic fluid with power-law equation of state $P
\propto \rho^\gamma$ (e.g., de Jong et al.\ 1980; Maloney 1988;
Elmegreen 1991; \VS, Passot \& Pouquet 1996); we use the term 
``effective polytropic exponent'' to refer to $\gamma$.\footnote{$^1$}{We
remark that the present effective polytropic exponent, which describes
the thermal behavior of gases with local heating and cooling in
equilibrium, is different from the adiabatic exponent $c_p/c_v$ (describing
thermal behavior with no heating or cooling) and that its usage does not imply
in any form that the system is in hydrostatic equilibrium. Also, note
that strictly speaking the polytropic equation is not properly an
equation of state, since the ideal-gas equation of state is also
satisfied at all times.}
In general, $\gamma$ is different from 1 for the global
ISM (Myers 1978), and possibly even for molecular clouds (de Jong et
al.\ 1980; Scalo et al.\ 1998). Note, however, that, since cooling
laws are often approximated by piecewise power-laws, the effective
polytropic exponent is also approximately piecewise
constant. Numerical simulations explicitly including heating and
cooling appropriate for atomic gas indeed show that the corresponding
rates are faster than the dynamical rates by factors of at least 50
(Passot et al.\ 1995; \VS\ et al.\ 1996), confirming earlier
estimations (Spitzer \& Savedoff 1950; Elmegreen 1993b). However,
those simulations
depart from pure power-law equations of state near star formation
sites.  For optically thin gas with rapid heating and cooling, an
effective polytropic equation of state with parameterized $\gamma$ may
therefore be considered as the next level of refinement over
isothermal or adiabatic laws; from the above simulations, $\gamma$ is
found to take values between 0 and 1/2 in the warm and cold
``phases''. For molecular material which is optically thick and
turbulent, it is not yet clear from simulations how well an isothermal
or effective polytropic law may be satisfied.

The choice of cooling and heating functions in the above-mentioned
simulations implies the absence of an isobaric thermal instability
(Field, Goldsmith \& Habing 1969) at densities $\sim 0.5$--5 cm$^{-3}$
and temperatures between $10^2$ and $10^4$ K.  There exist realistic
values of interstellar parameters, such as the background UV field,
coolant depletion and grain abundances, for which such an instability
is indeed not realized (e.g., Draine 1978; Wolfire et al.\ 1995).
This choice of parameters has allowed the analysis of the role of
turbulence separately from that of the thermal instability. Even
though it is desirable to study these processes in combination, the
thermal instability is unlikely to be the main driver of interstellar
motions, since thermal pressure is smaller, by at least a factor of 3,
than each of the other principal forms of interstellar pressure
(turbulent, magnetic and cosmic ray), and at least $\sim 10$ times
smaller than their sum (e.g., Boulares \& Cox 1990).

The stability of fluid parcels compressed in $n$ dimensions by shocks
or turbulence requires $\gamma > \gamma_{\rm cr} \equiv 2(1-1/n)$
(Chandrasekhar 1961; McKee et al.\ 1993; \VS\ et al.\ 1996). Shocks
are likely to have $n=1$, but generic turbulent compressions can have
any dimensionality $n\leq 3$. The latter authors have also shown
numerically the production of small-scale collapsing regions within a
globally stable turbulent medium, both in magnetic and non-magnetic
cases, although the former still requires the collapsing region to be
super-critical (cf. \S IV.B). They also noted that turbulence-induced
collapse generally involves small fractions of the mass available in
the flow, in contrast to the case of globally unstable regions, and
showed that the density jump $\rho_2/\rho_1 \equiv X$ across shocks in
a polytropic medium 
satisfies $X^{1+\gamma} - (1+ \gamma M_s^2)X + \gamma M_s^2=0$. In
particular, this recovers the well known result that $X=M_s^2$ for
$\gamma=1$, but also implies that in the limit of vanishing $\gamma$,
$X \rightarrow \exp(M_s^2)$.  An important consequence of this
effective polytropic behavior is that, if $0 < \gamma <1$, denser
regions are {\it colder}. Upon the production of turbulent density
fluctuations, the flow develops a temperature distribution similar to
that resulting from isobaric thermal instabilities ($\gamma < 0$), but
without the need for them. Essentially, turbulent ram pressure
provides the drive which in the thermally unstable case is provided by
thermal pressure. Note, however, that in the turbulent $\gamma >0$
case there are no sharp phase transitions. In this scenario, the
near-constancy of the thermal pressure among the various
moderate-density phases of the ISM is an incidental consequence of a
small value of $\gamma$ rather than a cloud confining agent
(Ballesteros-Paredes, \VS\ \& Scalo, 1998, hereafter BVS98). The
increase in thermal pressure in molecular clouds is a consequence of
$\gamma$ being closer to unity at the relevant densities (Scalo et
al.\ 1998), although the thermal pressure is still subdominant with
respect to other sources of pressure for scales $\gtrsim 0.1$ pc.

For three-dimensional compressions, the minimum Mach
number $M_0$ necessary to induce collapse by the velocity field has
been computed by several authors as a function of $\gamma$ and the
mass $m$ of the cloud in units of the Jeans mass. It is found that
$M_0 \propto \ln m$ for the isothermal ($\gamma=1$) case (Hunter
1979), $M_0 \propto m^{\gamma-1/(4-3\gamma)}$ for $4/3 > \gamma >1$
(Hunter \& Fleck 1982) and $M_0 \geq \sqrt{10/3(1-\gamma)}$ for
$0 < \gamma < 1$ (Tohline
et al.\ 1987). Note that the latter result is independent of the
cloud's mass, at least for perfect spherical geometries.

Recently, BVS98 have investigated some implications of the scenario
that clouds are turbulent density fluctuations by exploring the
properties of clouds formed in 2D numerical simulations of the ISM
(Passot et al.\ 1995) on scales between 1.25 pc and 1 kpc, including
standard atomic-gas cooling rates, diffuse background heating, modeled
stellar ionization heating, self-gravity, the Coriolis force, galactic
shear and magnetic fields. It should be stressed that the clouds {\it
form} during the evolution of the simulations, rather than being
started with some pre-defined conditions, and have estimated
lifetimes (BVS98) comparable to observational estimates (e.g., Bash et
al.\ 1977; Blitz \& Shu 1980; Blitz 1994). BVS98 find that the
velocity field is in 
general continuous across the boundaries, with kinetic surface terms
in the Eulerian form of the Virial Theorem (McKee \& Zweibel 1992) of
comparable magnitude to the total kinetic energy contained within the
clouds.

BVS98 also suggest that the formation of hydrostatic structures within
a turbulent medium is not possible unless the effective polytropic
exponent $\gamma$ increases during the process of collapse initiated
by a turbulent compression. Such a change in $\gamma$ will not occur
until protostellar densities are reached if thermal pressure alone is
considered, so in general the production of hydrostatic structures
appears unlikely.

\mainsection{{I}{I}{I}. CLOUD STRUCTURE}
\backup
\subsection{A.~~Scaling Relations and Energy Spectra}

One of the most conspicuous properties of molecular clouds (shared
with diffuse HI clouds), are the so-called Larson (1981) relations,
which in their most common form read $\Delta v \propto R^{1/2}$ and
$\rho \propto R^{-1}$, where $R$ is a characteristic spatial scale
(e.g. cloud size), $\Delta v$ is a characteristic velocity difference
(e.g. linewidth or line centroid difference), and $\rho$ the cloud's
mean density.\footnote{$^1$}{Note that Larson's original exponents
were slightly different, being 0.38 for the linewidth-size relation
and $-1.1$ for the density-size relation.} The most generally accepted
explanation of the origin of these relations, as stated in the
Introduction, is that the clouds are in virial equilibrium between
gravity and turbulent support (which causes one of the two relations
to become a consequence of the other), together with some other
assumption, such as that the clouds are magnetically critical and that
the magnetic field does not vary much from one region to another
(e.g., Shu et al.\ 1987; Mouschovias 1987), which fixes the other
relation.  However, molecular clouds may not always be gravitationally
bound (Blitz 1994); indeed, except at the largest scales, observed
internal clumps within clouds are {\it not} gravitationally confined
(Bertoldi and McKee 1992).  Being highly dynamic entities, even bound
clouds may generally not be in {\it static} virial equilibrium
(Ballesteros-Paredes \& \VS\ 1997).

Alternative explanations under turbulent conditions have been proposed
as well. Kolmogorov-like arguments based on, for example, cascades of
angular momentum (Henriksen \& Turner 1984) or kinetic energy density
(Ferrini et al.\ 1983; Fleck 1996), are reviewed in \VS\ (1998). Here
we briefly discuss results from numerical simulations.

As mentioned in \S\ I.B, for highly compressible regimes, the
turbulent energy spectrum is expected to approach the form $E(k)
\propto k^{-2}$ of shock-dominated Burgers turbulence.  If we assume
that the observed linewidths measure the mean square turbulent
velocity $u_l$ over scales smaller than $l$, we can write $u_l^2 = 2
\int_{2 \pi/l}^\infty E(k) dk \propto \int_{2 \pi/l}^\infty k^{-2} dk
\propto l$, so that the observed velocity dispersion-size relation
appears to emerge naturally.  However, the identification
of $\Delta v(l)$ with $u_l$ is not trivial. While $\Delta v(l)$ is the
linewidth within a beam of width $l$ integrated over the whole line of
sight, $u_l$ is an average over an idealized ensemble (and, in
practice, over all space) for volumes of size $l^3$.  Thus, the
identification of $u_l$ with $\Delta v(l)$ may depend on each beam
being dominated by a single component of scale $\sim l^3$;
hierarchical density fluctuations could potentially produce the
required structure.  The data of Issa et al. (1990) and Falgarone et
al.\ (1992), which include positions in the sky away from brightness
maxima and still exhibit a similar relation (slope $\sim 0.4$), seem
to support the turbulent origin of the $\Delta v$--$R$ relationship
(Issa et al.\ also considered random 
positions in their maps). Peng et al.\ (1998) have recently studied a
sample of CCN clumps, most of which are reported to be gravitationally
unbound, yet seem to follow Larson's linewidth-size relation as well.
However, Bertoldi \& McKee (1992) found that the smaller, non-
self-gravitating clumps in their large study sample did not follow
Larson's linewidth-size relation.

\VS, Ballesteros-Paredes \& Rodr\'iguez (1997, hereafter VBR97) have
surveyed the clouds appearing in the 2D numerical simulations of
Passot et al.\ (1995), finding a velocity dispersion-to-size relation
with a logarithmic slope $\sim 0.4$ as well (albeit with a large
scatter).  The energy spectrum of those simulations is indeed of the
form $k^{-2}$. To study the development of the energy spectrum, 
simulations of MHD turbulence in which driving is localized in  
wavenumber space at scales smaller than the box have been performed in
slab geometry (``1 2/3 D'') (Gammie \&\ Ostriker 1996), and
fully 3D geometry 
(Stone 1998; Stone, Ostriker, \& Gammie 1998).  For both cases,
extended spectra develop in $k$-space above and below the range of
driving frequencies, indicating the development of both ``direct'' and
``inverse'' cascades.  The 1 2/3 D models, which were evolved over
very long times, show spectral slopes $-2$ or slightly steeper both
above and below the forcing scale.  The 3D models, although more
limited in dynamic range, show combined $u_k^2 +B_k^2$ slopes between
$-2$ and $-5/3$ for $k$ larger than the forcing scale, with the former
(latter) occurring in weaker (stronger) magnetic field models.  The
slope of $u_k^2$ alone is $-2$ (stronger fields) or slightly steeper
(weaker fields).

Overall, turbulent MHD simulations tend to evolve toward global energy
equipartition. For example, Ostriker et al.\ (1998) and Stone et al.\
(1998) find perturbed magnetic energies between 30-60\% of kinetic
energies over a wide range of $\beta$. Kinetic and magnetic spectra
are within factors of a few from each other (Passot et al.\ 1995),
implying equipartition at all scales.

VBR97 also find cloud mass spectra of the
form $dN(M)/dM \propto M^{-1.44\pm0.1}$, consistent with the low end
of observational estimates (e.g., Blitz 1993), suggesting that the
simulations reproduce well a number of observational cloud properties,
even though they are two-dimensional. Additionally,
VBR97 find that Larson's density-size relation is not satisfied in
general, but rather seems to be the upper envelope in a $\log
\rho$-$\log R$ diagram of the clouds' locus, i.e., it is satisfied
only by the densest clouds at a given size (implying largest column
densities).  Low-column density clouds would naturally escape
observational surveys utilizing limited amounts of integration time
(Larson 1981; Kegel 1989; Scalo 1990), suggesting that this relation
may be an observational artifact. The off-peak data of Falgarone et
al.\ (1992) are consistent with this suggestion.  In summary, these
results can be interpreted as implying that a $\Delta v$-$R$ relation
comparable to the observed scalings may be established globally as a
consequence of the development of compressible MHD turbulence, while
the density-size relation may occur only for gravitationally bound
clouds within this turbulent field. Further work is necessary to
confirm this possibility.

\subsection{B.~~Turbulent Pressure}

\VS, Cant\'o \& Lizano (1998) have performed 2D and 3D numerical
simulations of isothermal gravitational collapse in initially
turbulent clouds, following the evolution of the velocity dispersion
as the mean density increases during the collapse.  They found
power-law behavior of the form $P_t \sim \rho^{\gamma_t}$ for the
``turbulent pressure''. In particular, for slowly-collapsing magnetic
simulations, $\gamma_t \sim 3/2$, consistent with the result of McKee
\& Zweibel (1995) for the adiabatic exponent of Alfv\'en waves upon
slow compression. However, non-magnetic and rapidly-collapsing
(shorter Jeans length) magnetic simulations have $\gamma_t \sim 2$.
Gammie \&\ Ostriker (1996) also verified the McKee \&\ Zweibel
scalings $P_{wave}\propto \rho^{3/2}$ and $P_{wave}\propto \rho^{1/2}$
for Alfv\'en wave pressure of, respectively, an adiabatically
contracting medium, and propagation along a density gradient.  For
freely-evolving decay simulations, however, they found only a weak,
variable correlation between perturbed magnetic pressure and density.
These results are incompatible with the so-called ``logatropic''
equation $P \sim \ln \rho$, warning against its usage in 
dynamical situations.

\subsection{C.~~The Density Field}

The turbulent density field has a number of relevant statistical and
physical properties. Besides the density-size scaling relation (Sec.\
III.A above), which may or may not be a true property of clouds, these
exhibit hierarchical nesting (e.g., Scalo 1985; Houlahan \& Scalo
1990, 1992; \VS\ 1994) and evidence for fractal (e.g., Falgarone et
al.\ 1991) or even multi-fractal structure (Chappell \& Scalo
1998). The spatial and statistical distribution of the density field
is crucial in the study of star formation and the understanding of the
stellar Initial Mass Function (IMF).

In order to construct a real theory of the IMF (or at least of the
cloud mass spectrum, if the actual masses of stars turn out to be
rather independent of their parent clumps) it is necessary to have a
complete knowledge of the density statistics. A simple theory based
exclusively on the probability density function (pdf, also commonly
referred to as the density distribution function) of the fluid density
(Padoan 1995; Padoan, Nordlund \& Jones 1997) has been criticized by
Scalo et al.\ (1998) who point out that, in addition to the
probability of occurrence of high density sites, it is also necessary
to know how much mass is contained in these fluctuations; this
requires information on multi-point statistics. 

As a first step
towards the understanding of density fluctuations in the ISM, the pdf
of the density in polytropic gas dynamics ($P\propto \rho^\gamma$) has
been investigated as a function of the rms Mach number, $\gamma$,
and the mean magnetic field strength (\VS\ 1994;
Scalo et al.\ 1998; Passot \& \VS\ 1998; Nordlund \& Padoan 1998;
Ostriker, Gammie, and Stone 1998; Stone et al. 1998).
In the isothermal case ($\gamma=1$) the density pdf is close to a
lognormal distribution for every value of the Mach number.  For
polytropic cases with $\gamma <1$ or $\gamma>1$, a power law develops at 
densities larger than the mean or smaller than the mean, respectively, 
the effect being
enhanced as the Mach number increases. This behavior is a consequence
of the dependence of the local Mach number with the density in each
case (Passot \& \VS\ 1998; Nordlund \& Padoan 1998). It has been
verified in one-dimensional numerical simulations of a forced
polytropic gas, but appears to be supported also in several dimensions
and in the presence of thermal heating and cooling which yield
``effective'' polytropic behavior (Scalo et al.\ 1998). In the limit
of high Mach numbers and/or vanishing effective polytropic exponent,
the behavior does not coincide with that of the Burgers equation
(Passot \& \VS\ 1998). One implication
of these results is that it should be possible in principle to
determine the actual effective $\gamma$ of the medium from
observational determinations of its pdf, provided the problem of
deconvolving the projected pdf is solved and $\gamma$ does not vary
much in the observed region. Also, for example, the observation by
Scalo et al.\ (1998) that full simulations of the ISM including the
turbulent magnetic field have pdfs consistent with $\gamma < 1$
suggests that the MHD waves do not give appreciably large values of
$\gamma$ (as is also suggested by the $P\propto \rho^{1/2}$ scaling of 
propagating linear-amplitude waves; cf. McKee \& Zweibel 1995).

The mean mass-averaged value of $\log(\rho/\bar\rho)$ increases with
the Mach number $M_s$; for isothermal models in 2.5D and 3D
respectively, Ostriker et al. (1998) and Padoan, Jones \&\ Nordlund
(1997) both find a logarithmic dependence on $M_s^2$. In large-Mach
number, high-resolution 1D isothermal models by Passot \& \VS\ (1998),
a linear dependence is found between the variance of the density
logarithm and $M_s^2$, leading to a mass-averaged value of $\log
\rho/\bar\rho$ that varies like $M_s^2$; this difference with
higher-dimensional simulations (linear vs. logarithmic scaling with
$M_s^2$) likely arises because the 1D simulations have a purely
compressive velocity field and the higher-dimension simulations do
not.  Ostriker et al.\ (1998) and Stone et al.\ (1998) show that the
largest mean contrasts in the density logarithm occur in models with
the strongest mean magnetic fields (see also Pouquet, Passot and
L\'eorat 1990).

The spectrum of the density field has recently started to be
investigated. Padoan, Jones \&\ Nordlund (1997) have reported a steep
logarithmic slope $\sim -2.6 \pm 0.5$ for low-resolution 3D
simulations of isothermal turbulence. Scalo et al.\ (1998) have
reported two regimes, one with a slope $\sim -0.9$ at low $k$ and
another with slope $\sim -2.4$ at large $k$ for high-resolution 2D
simulations of the ISM with heating and cooling. The steeper slopes
may be due to inadequate resolution. On the other hand, Lazarian
(1995) has produced an algorithm for deconvolving projected HI
interferometric spectra, favoring a slope $\sim -1$. Equivalent work
for molecular clouds and high-resolution simulations in the
corresponding regimes are necessary to resolve this issue.

An interesting application has been given by Padoan, Jones \& Nordlund
(1997), who have shown that a turbulent density field with a power-law
spectrum (though steep) and a lognormal pdf produces simulated plots
of extinction dispersion vs.\ mean extinction that compare well with
the analogous observational diagram for an extinction map of a dark
cloud.  
Padoan \& Nordlund (1998a) also suggest that models with weaker mean
magnetic fields (such that $M_A\sim 10$) have extinction dispersion
vs.\ mean extinction plots that show better agreement with observations than
do models with stronger mean magnetic fields (such that $M_A\sim 1$);
however, these models do not include self-gravity, which would affect
the distribution of column densities (i.e. extinctions).

\subsection{D.~~Correlations Among Variables}

Numerical simulations are especially useful in the investigation of
the spatial correlation among physical variables, a crucial ingredient
in the formation of stars. Of particular interest is the correlation
of magnetic field strength with density, as well as the correlation
between field direction and the topology of density features.

The topology of the clouds is extremely clumpy and filamentary -- for
an example, see Figure 1.  In general, the magnetic field exhibits a
morphology indicative of significant distortion by the turbulent
motions, with greater magnetic field tangling in cases with weak mean
magnetic fields (Ostriker et al.\ 1998, Stone et al.\ 1998).  Magnetic
fields may be either aligned with or perpendicular to density features
(cf. Figure 1; Passot et al.\ 1995; \VS\ \& Passot 1998; Ostriker et al.\
1998; Stone et al.\ 1998), with the former trend most visible at the
boundaries of supershells (Gazol \& Passot 1998a), and the latter most
prominent for strong-field simulations in which field kinks and
density maxima coincide (Gammie \&\ Ostriker 1996; Ostriker et al.\
1998).

Globally, no clear trend between density and magnetic field intensity
is found in the simulations, but very weak correlations are observed
(Passot et al.\ 1995, Gammie \&\ Ostriker 1996).  For example, in the
large-scale 2D ISM models of Passot et al.\ (1995), the field strength
$B$ varies from $\sim 10^{-2}\ \mu G$ in the low density intercloud
medium to $\sim 25\ \mu G$ in the densest clouds ($n \sim 50$
cm$^{-3}$, size $\sim$ several tens of pc), although vanishing field
strengths are also found in those regions.  In 3D simulations with
weak mean magnetic fields ($B_0=1\mu$G), Padoan \& Nordlund (1998a)
found a large dispersion in the values of $B$ as a function of density
$n$, but a power-law correlation $B\propto n^{0.4}$ in the upper
envelope of this distribution.  In stronger-mean-field simulations
($B_0=30\mu$G), they found little variation of $B$ with $n$.  Padoan
\& Nordlund argue that the $B-n$ upper-envelope correlation found in
the former case supports the notion that the mean fields in molecular
clouds are weak, because a similar $B-n$ scaling has been observed for
measured Zeeman field strengths (e.g. Crutcher 1998).  However,
because the densities over which the envelope correlation is found in
simulations is much smaller than the density regime in which a
power-law $B-n$ relation is observed in real clouds, and because the
simulations do not include gravity, the conclusion remains
controversial.

To explore how line spectra may vary spatially in simulated clouds, Padoan
et al.\ (1998) have produced synthetic non-LTE spectra of various
molecular transitions from models with weak and strong mean magnetic
fields.  They find that their super-Alfv\'enic model reproduces the
observational trend of line-width vs.\ integrated temperature found by
Heyer, Carpenter \& Ladd (1996), while the equipartition model gives a
weaker trend. However, these models do not include self-gravity, which
would affect the density distributions and density-velocity
correlations.  Thus, it remains uncertain whether the weak-mean-field
model of molecular clouds advanced by Padoan \& Nordlund (1998a) truly
provides a better fit to observations. 
 
\mainsection{IV.~~CLOUD EVOLUTION}

\subsection{A.~~Dissipation Rates and Turbulence Maintenance}

As described above, the first observations of supersonic internal
cloud velocities immediately led to the questions -- which have
persisted up to the present -- of what creates these large-amplitude
motions, and how they are maintained.  Since the large-scale ISM is
itself turbulent, turbulent motions may be incorporated into cold
clouds from their formation stages, being part of the same
continuum. In addition, there may be ongoing inputs which tap energy
from larger scales in the Galaxy, or from smaller scales within the
cloud, particularly associated with various aspects of star formation
(see e.g. Scalo 1987; Miesch and Bally 1994; Kornreich \& Scalo 1998).
One of the first steps in understanding cloud evolution, and relating
this evolution to the initiation of star formation and dynamical
feedback, is to assess the rate of turbulent decay under the range of
conditions representative of dark clouds and GMCs.

 From dimensional analysis, the decay rate of kinetic energy per unit
mass must scale as $\dot E\sim v^3/R$, where $v$ is some
characteristic speed (or weighted product of two or more speeds), and
$R$ is some characteristic scale.  In incompressible turbulence
(cf. \S I.B), the only characteristic scales are that of the box
($L$), that of the flow velocity difference $u_L$ on the largest
scale, and the value of the small-scale viscosity $\nu$.  The velocity
dispersion over the whole box $\sigma_v\sim u_L$ provided the spectral
slope is $-1$ or steeper.  The first two quantities determine the
decay rate, $\sigma_v^3/L$, while the last sets the spatial
dissipation scale to $l_{min}/L\sim (\nu/L\sigma_v)^{3/4}\equiv
Re^{-3/4}$.  For a nearly pressureless (i.e. highly compressible)
fluid, again possessing the same three characteristic scales, the same
dissipation rate would apply, except that energy would be transferred
by a shock directly from the largest scale to the dissipation scale
within a flow crossing time.  For a magnetized flow of finite
compressibility, on the other hand, other velocity scales in addition
to $\sigma_v$ -- namely those associated with thermal pressure and
magnetic stress, $c$ and $v_A$ -- enter the problem, and may
potentially influence the scaling of the dissipation rate.

In molecular clouds, the thermal pressure is very low, and therefore
strong dissipation in shocks is expected.  However, from the time of
early observations until quite recently, it has widely been considered
likely that the ``cushioning'' effect of magnetic fields would
significantly reduce kinetic energy dissipation for motions transverse
to the mean field, provided that turbulent velocities remain
sub-Alfv\'enic.  Low dimension (1 2/3D) simulations (Gammie \&\
Ostriker 1996) provided some support for this idea, in that they found
a scaling of dissipation rate with $\beta$ in quasi-steady-state as
$\dot E \propto \beta^{1/4}$, such that magnetic fields of a few tens
of $\mu G$ could potentially provide a factor $\sim 3$ reduction in
dissipation compared to that in weak-field ($\beta=1$) cases.
However, very recent higher-dimension numerical simulations of both
forced and decaying turbulence have shown that although some
differences remain between weak-field and strong-field cases,
dissipation rates are never substantially lower than the predictions
of unmagnetized turbulence.  Below we describe specific results.

The recent 3D MHD decay simulations (Mac Low et al.\ 1998; Stone et
al.\ 1998; Padoan \&\ Nordlund 1998a) have followed the evolution of
Mach-5 turbulence with a variety of initial velocity spectra and
$\beta$ ranging from $0.02$ to 2, and compared to unmagnetized models.
These models have uniform initial $\bf B$.  Stone et al.\ (1998) have
also included simulations of decay from fully-saturated turbulence,
and from saturated turbulence with initial density fluctuations
suppressed.  These experiments all show kinetic energy decay times
(defined as the time for kinetic energy reduction by 50\%) in the
range $\sim 0.4-1 t_f$, where $t_f$ is the flow crossing time $l/u_l$
on the main energy-containing scale.  In Stone et al.\ (1998), the
difference in decay time between the strongest-field case
($\beta=0.02$, corresponding to $44\mu G$ for $n_{H_2}=10^3{\rm
cm}^{-3}$ and $T=10$K) and the unmagnetized run is less than a factor
of two.  In decay models, Mac Low et al.\ (1998) and Stone et al.\
(1998) find late-time power law dependence of the turbulent energy as
$E\propto t^{-\eta}$ with $\eta=0.8-1$.

To assess turbulent decay rates for a quasi-steady
state, Stone et al.\ (1998) performed simulations in which a fixed
mechanical power $\dot E_K$ is input to the flow in the form of
random, uncorrelated velocity perturbations. A saturated state with
energy level $E_K$ is reached after $\sim t_f$, with only relatively
small differences in the saturation energy between magnetized and
unmagnetized models; the turbulence dissipation times $E_K/\dot E_K$
range between $0.5-0.7 t_f$.  Mac Low (1998) has found similar results,
and suggests that the power-law time dependence of decay models arises
from a secular increase in the smallest scale of turbulence.  The
implications of the short turbulent dissipation time for potential
cloud support are discussed below in \S IV. B.

Kornreich \& Scalo (1998) have recently considered the problem of
turbulence maintenance in molecular clouds, comparing the average time
between external shock wave passages through a cloud with the energy
decay time, finding that they are comparable and implying that this
``shock pump'' is capable of sustaining the cloud turbulence. An
additional result is the proposal of a cascade-like mechanism for the
compressible case, in which vorticity is generated behind shocks,
which in turn rapidly produces new smaller-scale shocks, and so
on. This is due to an interesting asymmetry of the evolution equations
for the vorticity and divergence of the velocity field (for recent
discussions, see \VS\ et al.\ 1996; Kornreich \& Scalo 1998): the
nonlinear transfer will produce compressible modes out of purely
rotational ones, but the converse is not true. This can also be
understood in terms of the well-known Kelvin theorem of conservation of
circulation.

The mechanisms of vorticity production have also been investigated
numerically. Simulations of compressible turbulence with purely
potential forcing indicate that a negligible amount of kinetic energy
is transferred from compressible to solenoidal modes (Kida \& Orszag 1990;
\VS\ et al.\ 1996). Vorticity generation behind curved shocks or shock
intersections, as well as the vortex stretching term do not seem to
be efficient processes to maintain a non-negligible level of vorticity
in the flow. However, in the presence of the Coriolis force, thermal heating
(through the baroclinic term) or magnetic field, near equipartition
between solenoidal and compressible modes is easily obtained (\VS\ et
al.\ 1996).

\subsection{B.~~Turbulence and Cloud Support Against Gravity}

An issue that is often discussed in tandem with turbulent dissipation
is the question of cloud support against self-gravity.  Cold, dark
clouds and GMCs have typical Jeans lengths $\sim 2$ pc, such that the
whole cloud entities exceed the Jeans mass $M_J=\rho L_J^3$ by factors
of a thousand or much more.  This is just another way of stating that
thermal pressure gradients would be powerless to prevent
self-gravitating runaway.  If not for the intervention of other
dynamical processes, wholesale cloud collapse would be the rule.

``Turbulent pressure'' is often invoked as a means to counter gravity.
For a weakly compressible medium, the $\rho v^2$ Reynolds stresses are
naturally associated with this turbulent pressure.  
The effect of turbulence on the gravitational instability in absence
of magnetic field was first investigated by Chandrasekhar (1951) who
suggested an increase of the Jeans length and later by Bonazzola et
al.\ (1987, 1992) and \VS\ \& Gazol (1995) who predicted a reversal of the
Jeans criterion when certain conditions on the energy spectrum are met.

In a strongly compressible and radiative medium, the collisions of 
supersonically converging streams of gas are nearly inelastic, so the
Reynolds stress does not itself act as an effective pressure.
Turbulent motions can, however, generate turbulent magnetic fields
(and vice versa); these fluctuating magnetic fields exert pressure and
tension forces on the medium.  Several authors (Shu, Adams, \& Lizano
1987; Fatuzzo and Adams 1993; McKee and Zweibel 1995) have suggested
that, in particular, the time-dependent magnetic field perturbations
associated with Alfv\'en waves could potentially be important in
providing ``wave pressure'' support against gravity along the mean
magnetic field direction in a cloud.  This mean-field axis is the most
susceptible to collapse; in the orthogonal directions, magnetic
pressure suppresses gravitational instability in a homogeneous cloud
as long as $n_J\equiv L/L_J<(\beta/2)^{-1/2}$ (Chandrasekhar and Fermi
1953), and suppresses instability in a cloud pancake of surface
density $\Sigma$ provided $\Sigma/B<1/(2\pi\sqrt{G})$ (cf. Mouschovias
and Spitzer 1976; Tomisaka, Ikeuchi, and Nakamura 1988).

An exact derivation of the Jeans criterion is hardly possible in the
MHD turbulent case. An attempt in this direction has been made by
considering the linear stability of a self-gravitating medium, permeated
by a uniform magnetic field $B_0$ along which a finite amplitude
circularly polarized Alfv\'en wave propagates. For perturbations along
$B_0$ it is found that 
the Alfv\'en wave increases the critical Jeans length (Lou 1997).
In the case of perturbations perpendicular to the mean field
Gazol \& Passot (1998b) have shown that the medium is less stable
in presence of a moderate amplitude  Alfv\'en wave. For large amplitude
waves however, McKee \& Zweibel (1995) show that the waves
have an isotropic stabilizing effect.

Gammie \&\ Ostriker (1996) verified, using simulations in
1 2/3 D, that Alfv\'en waves of
sufficient amplitude (such that $n_J\lesssim \delta v_A/2c_s$) can
indeed prevent collapse of slab-clouds along the mean field direction.
These simulations included both cases with decaying and quasi-steady
forced turbulence.  For 1D decay models, the turbulent dissipation
rate is low enough such that clouds with initial turbulent energy
above the limit remain uncollapsed for times up to $t_g\equiv L_J/c$
($\sim 10$ Myr for typical conditions). However, more recent
simulations performed in higher dimensions have shown that the greater
dissipation rates quench turbulence too rapidly for magnetic
fluctuations to prevent mean-field collapse.  
Because 3D dissipation times for both magnetized and
unmagnetized flows are smaller than the flow crossing
time $t_f$ (cf. \S IV A), they will also be less than the
gravitational collapse times $\approx 0.3 t_g$ for clouds that are
virialized (e.g. observed cloud scalings yield $t_f\sim 0.5 t_g$); thus
cloud support cannot be expected for un-regenerated turbulence.
Self-gravitating, magnetized 2 1/2D simulations of Ostriker et al.\
(1998) have already demonstrated this result directly, and concluded 
that without ongoing energy inputs, only the strength of the mean
magnetic field is important in determining whether or not clouds
collapse in times $< 10 $Myr.  Numerical simulations have also been
performed which allow for ongoing turbulent excitation.  In
unmagnetized 2D models, L\'eorat et al.\ (1990) showed that
large-scale gravitational collapse can be prevented indefinitely
provided a high enough Mach number is maintained by forcing, and the
energy is injected at small enough scales. For magnetized models,
Gammie \&\ Ostriker (1996) found similar results for 1 2/3 D
simulations (as well as for unpublished 2D simulations).

Ballesteros-Paredes \& \VS\ (1997) have measured the overall virial
balance of clouds in a ``survey'' of the 2D high resolution
simulations of the ISM of Passot et al.\ (1995). It was
found that the gravitational term appearing in the virial theorem
is comparable to the sum of the other virial terms for the largest
clouds, but progressively loses importance on the average as smaller
clouds are considered. Nevertheless, the scatter about this average
trend is large, and a small fraction of small clouds have very large
gravitational terms which overwhelm the others and induce
collapse. In this scenario, the low efficiency of star formation is
understood as a consequence of the intermittency of the turbulence.

\subsection{C.~~Evolution of the Star Formation Rate}

Numerical simulations on the large scales provide information on the
star formation (SF) history as well (\VS\ et al. 1995; Gazol \& Passot
1998a). In these models, stars form whenever a certain density threshold is
exceeded, rather than being treated as a separate fluid (e.g., Chiang \&
Prendergast 1985; Rosen, Bregman \& Norman 1993). A nontrivial result is the
development of a self-sustained cycle of SF in which the turbulence,
aided by self-gravity,
contains enough power to produce star-forming
clouds, while the energy injected by the
stars is sufficient to regenerate the turbulence. Due to the strong
nonlinearity of the SF scheme, this cycle is highly
chaotic and particularly intermittent in the presence of
SN. Self-propagating SF in supershells is very efficient, increasing the
SF rate in active periods, but the destructive power of superbubbles
is so large that the system requires a longer time to create new SF
sites once the shells have dispersed. 
A similar result was obtained by V\'azquez \& Scalo (1989) in a simple
model for gas-accreting galaxies. This behavior is consistent with recent
observations suggesting that the SF history in galaxies is highly
irregular (e.g., Grebel 1998). Another interesting point concerns
the influence of the strength of the uniform component of the magnetic
field $B_0$ (Gazol \& Passot 1998a). The star formation rate is found
to increase with $B_0$ (with $B_0=5 \mu G$ it is larger by a factor
$\sim 3$ compared with the case $B_0=0$), as long as $B_0$ is not too
large. For very large $B_0$, it decreases due to the rigidification of
the medium.

\mainsection{V.~~CONCLUSIONS} 

In this chapter we have reviewed a vast body of results which address
the problem of MHD compressible (MHDC) turbulence and its implications
for problems of cloud and star formation. Most of these results are
new, appearing after the last Protostars \& Planets Conference.
Interstellar turbulence is inherently a multi-scale and
non-equilibrium phenomenon, and thus appears to play a fundamental
role in the formation, evolution and determination of cloud structural
properties. Most of the studies reviewed here have relied on direct
numerical simulations of MHDC turbulence in a variety of regimes,
ranging from isothermal to polytropic to fully thermodynamic, the
latter including parameterized heating and cooling and modeled star
formation. High-resolution 2D and 3D simulations of the ISM at ``box''
scales from 1 kpc down to a few pc have shown that the clouds formed
in them reproduce well a number of observational cloud properties
(such as clumpy structure and linewidth-size scalings), while
suggesting that some others (e.g. Larson's (1981) density-size
relation) may either arise from selection effects or require other
special conditions.  Simulations to date have shown that the old paradigm that
MHDC turbulence should dissipate much more slowly than non-magnetic
turbulence may be incorrect, bringing back the necessity of strong
large-scale fields for magnetic cloud support and/or of continued
energy injection in order to sustain the turbulence.   The idea for
self-regulation of star formation by turbulent feedback remains
promising, and it is also possible that physical processes not 
yet incorporated in simulations may reduce the turbulent dissipation rate.

A large number of questions remain unanswered, however. At the larger
scales, it is necessary to understand the interplay between turbulence,
large-scale instabilities, and spiral waves in the formation of
molecular clouds and complexes, as well as to investigate the
processes that lead to cloud destruction.  Quantitative assessments
must be developed for the efficiency of turbulent excitation from
internal and external sources.  A theory of the IMF, or at least of
the cloud mass spectrum, requires the knowledge and understanding of
multi-point density statistics.  Better understanding of the parameter
dependence of these and other questions in cloud evolution and
structure will be required to address the problem of ``deriving'' the
star formation rate and efficiency, in answer to the oft-posed challenge
of extragalactic astronomers.  Comparison with
observations is crucial to discriminate among models, but will always
face the degeneracy limitations associated with projection effects.

Future research in these areas is likely to include more physical
processes (e.g., ambipolar diffusion, radiative transfer, and global
environment effects, such as the galactic spiral potential and
variations in external radiation with galactic position), to perform
direct statistical comparison with observations of molecular lines,
and polarization and IR measurements, and to work at higher
resolutions in 3D, if the multi-scale nature of the problem is to be
captured adequately.  With so much still untried, there is
considerable ground to cover before reaching the long-range goal of
synthesizing the results of disparate numerical simulations into a
coherent theory of the turbulent ISM.  Nevertheless, the important
advances since PPIII show the success of numerical methods in
answering many longstanding questions in the theory of interstellar
MHD turbulence, and the opportunity for great progress before the next
{\it Protostars and Planets}.

\bigskip\bigskip

\leftline{\bf Acknowledgements} We are grateful to C. F. McKee for
helpful comments on this manuscript and to J. Scalo for enlightening
discussions.  This work has been supported in part
by grants UNAM/DGAPA IN105295, CRAY/UNAM SC008397, NAG5380 from NASA,
and by the CNRS "PCMI" National Program.

\vfill\eject

\vfill\eject

\null

\vskip .5in
\centerline{\bf REFERENCES}
\vskip .25in

\ref{Arons, J. \& Max C. E. 1975. Hydromagnetic waves in molecular
clouds. \apjl 196:77--81}

\ref{Ballesteros-Paredes, J., \VS, E. \& Scalo, J. 
1998. Clouds as turbulent density fluctuations. Implications for 
pressure confinement and spectral line data interpretation. \apj in
press (BVS98)}

\ref{Balsara, D.S., Crutcher, R.M., \& Pouquet, A. 1997.  Numerical
Simulations of MHD Turbulence and Gravitational Collapse.
In {\refit Star Formation, Near and Far}, Eds. S.Holt and L. Mundy.
(Woodbury NY: AIP Press), pp 89--92}

\ref{Bash, F. N., Green, E. \& Peters, W. L., III 1977. The Galactic
density wave, molecular clouds, and star formation. \apj 217:464--472}

\ref{Bate, M. R., Clarke, C. J., \& McCaughrean, M. J. 1998. 
Interpreting the mean surface density of companions in star-forming
regions. \mnras 297:1163--1181}

\ref{Bertoldi, F., and McKee, C. 1992. Pressure-confined clumps in
magnetized molecular clouds. \apj 395:140--157}

\ref{Blitz, L. \& Shu, F. H. 1980. The origin and lifetime of giant
molecular cloud complexes. \apj 238:148-157}

\ref{Blitz, L. 1993. Giant molecular clouds. In {\refit Protostars and
Planets III}, eds. E. H. Levy \& J. I. Lunine (Tucson: University of
Arizona Press), pp. 125--161}

\ref{Blitz, L. 1994. Simple things we don't know about molecular
clouds. In {\refit The Cold Universe}, ed. Th. Montmerle, C. J. Lada,
I. F. Mirabel \& J. Tr\^an Thanh V\^an (Gif-sur-Yvette: Editions
Frontiers), pp. 99-106}

\ref{Boulares, A. \& Cox, D. P. 1990. Galactic hydrostatic equilibrium 
with magnetic tension and cosmic-ray diffusion. \apj 365:544--558}

\ref{Braun, R. 1998. Properties of atomic gas in spiral galaxies. In
{\refit Interstellar Turbulence, Proceedings of the 2nd Guillermo
    Haro Conference}, Eds. Franco, J. \& Carraminana, A.
   (Cambridge:Cambridge University Press), in press}

\ref{Burgers, J. M. 1974. The nonlinear diffusion equation. (Dordrech:
Reidel)}

\ref{Carr, J. S. 1987. A study of clumping in the Cepheus OB 3
molecular cloud. \apj 323:170--178} 

\ref{Chandrasekhar, S. 1951. The gravitational instability in an
infinite homogeneous turbulent medium. Proc. Royal Soc. London 210:26--29}

\ref{Chandrasekhar, S., and Fermi, E. 1953.  Problems of Gravitational
Instability in the Presence of a Magnetic Field. \apj 118:116-141}

\ref{Chandrasekhar, S. 1961. Hydrodynamic and Hydromagnetic
 Stability (Oxford: Clarendon Press)}

\ref{Chappell, D. \& Scalo, J. 1998. Multifractal Scaling, Geometrical
Diversity, and Hierarchical Structure in the Cool Interstellar
Medium. \apj, submitted}

\ref{Chiang, W.-H. \& Prendergast, K. H. 1985. Numerical study of a
two-fluid hydrodynamic model of the interstellar medium and population
I stars. \apj 297:507--530}

\ref{Crutcher, R. M., Troland, T. H., Goodman, A. A., Heiles, C.,
Kaz\`es, I., \& Myers, P. C. 1993. OH Zeeman observations of dark
clouds. \apj 407:175--184} 

\ref{Crutcher, R. M., Troland, T. H., Lazareff, B., Kaz\`es,
I. 1996. CN Zeeman Observations of Molecular Cloud Cores. \apj 456:217--224}

\ref{Crutcher, R.M. 1998. Observations of Magnetic Fields in Dense
  Clouds:  Implications for MHD Turbulence and Cloud Evolution. In
  {\refit Interstellar Turbulence, Proceedings of the 2nd Guillermo
    Haro Conference}, Eds. Franco, J. \& Carraminana, A.
   (Cambridge:Cambridge University Press) }

\ref{de Jong, T., Dalgarno, A., \& Boland, W. 1980. Hydrostatic models
of molecular clouds. I. Steady state models. \aa 91:68--84}

\ref{Draine, B. T. 1978. Photoelectric heating of interstellar
gas. \apjs 36:595--619}

\ref{Elmegreen, B. G. \& Lada, C. J. 1977. Sequential formation of
subgroups in OB associations. \apj, 214, 725--741}

\ref{Elmegreen, B. G. \& Elmegreen, D. M. 1978. Star formation in
shock-compressed layers. \apj 220:1051--1062}

\ref{Elmegreen, B. G. 1991. Cloud formation by combined instabilities
in galactic gas layers - Evidence for a Q threshold in the
fragmentation of shearing wavelets. \apj 378:139--156} 

\ref{Elmegreen, B. G. 1993a. In {\refit Protostars \& Planets
III}, eds. E. H. Levy \& J. I. Lunine (Tucson: University of Arizona
Press), pp.\ 97--124.}

\ref{Elmegreen, B. G. 1993b. Star formation at compressed interfaces
in turbulent self-gravitating clouds. \apjl 419:29--32}

\ref{Elmegreen, B. G. 1995. Star Formation on a Large Scale.  In
{\refit Molecular Clouds and Star Formation}, eds. C. Yuan and Junhan
You (Singapore: World Scientific), pp.\ 149--205}

\ref{Falgarone, E., Phillips, T. G., \& Walker, C. K. 1991. The edges
of molecular clouds: fractal boundaries and density structure. \apj
378:186--201}

\ref{Falgarone, E., Puget, J.-L., \& P\'erault, M. 1992. The
small-scale density and velocity structure of molecular clouds. \aa
257:715--730}

\ref{Falgarone, E., Panis, J.-F., Heithhausen, A., P\'erault, M.,
Stutzki, J., Puget, J.-L., \& Bensch, F. 1998. The IRAM key-project:
small-scale structure of pre-star-forming regions. I. Observational
results. \aa 331:669--696} 

\ref{Fatuzzo, M., and Adams, F.C. 1993. Magnetohydrodynamic wave
    propagation in one-dimensional nonhomogeneous, self-gravitating
    clouds \apj 412:146--159

\ref{Ferrini, F., Marchesoni, F., \& Vulpiani, A. 1983. A hierarchical
model for gravitational compressible turbulence. \apss 96:83--93}

\ref{Field, G. B., Goldsmith, D. W., \& Habing,
H. J. 1969. Cosmic ray heating of the interstellar gas. \apjl 155:149--154}

\ref{Fleck, R. C. 1981. On the generation and maintenance of
turbulence in the interstellar medium. \apjl 246:151--154}

\ref{Fleck, R.C. 1996. Scaling relations for the turbulent,
non- self- gravitating, neutral component of the interstellar
medium. \apj 458: 739--741}

\ref{Gammie, C.F. 1996. Linear Theory of Magnetized, Viscous,
Self-gravitating Gas Disks. \apj 462:725-731} 

\ref{Gammie, C.F. and Ostriker, E.C. 1996. Can Nonlinear Hydromagnetic
        Waves Support a Self-gravitating Cloud? \apj 466:814--830}

\ref{Gazol, A. \& Passot, T. 1998a. A turbulent model for the interstellar
medium. III. Stratification and supernova explosions. \apj, submitted}

\ref{Gazol, A. \& Passot, T. 1998b. Influence of Alfv\'en waves on the
gravitational instability. \aa, submitted}

\ref{Goldreich, P. \& Kwan, J. 1974. Molecular clouds. \apj 189:441--453}

\ref{Goldreich, P. and Sridhar, S. 1995.  Toward a theory of interstellar 
        turbulence. II: Strong Alfv\'enic turbulence. \apj 438:763--775}

\ref{Goodman, A.A., Barranco, J.A., 
        Wilner, D.J., and Heyer, M.H. 1998. Coherence in Dense Cores. II. The 
        Transition to Coherence. \apj, 504:223--}

\ref{Grebel, E. 1998. Star Formation Histories of Local Group Dwarf
Galaxies. In {\refit Dwarf Galaxies and Cosmology}, eds. T. X. Thuan, C.
  Balkowski, V. Cayatte, \& J. Tran Thanh Van (Gif-sur-Yvette: Editions
Frontiers), in press}

\ref{Heiles, C., Goodman, A.~A., McKee, C.~F., \& Zweibel,
E.~G. 1993.  Magnetic fields in star-forming regions:
observations.  In {\refit Protostars and Planets III}, ed. E.
H. Levy \& J. I. Lunine (Tucson: University of Arizona Press),
pp. 279--326} 

\ref{Henriksen, R. N., \& Turner, B. E. 1984. Star cloud
turbulence. \apj 287:200--297}

\ref{Heyer, M. H., Carpenter, J. M., \& Ladd, E. F. 1996. Giant
molecular cloud complexes with optical HII regions: $^{12}$CO and
$^{13}$CO observations and global cloud properties. \apj 463:630--641}

\ref{Houlahan, P. \& Scalo, J. 1990. Recognition and characterization
of hierarchical interstellar structure. I. Correlation function. \apjs
72:133--152}

\ref{Houlahan, P. \& Scalo, J. 1992. Recognition and characterization
of hierarchical interstellar structure. II. Structure tree
statistics. \apj 393:172--187}

\ref{Hunter, J. H., Jr. 1979. The influence of initial velocity fields
upon star formation. \apj 233:946--949}

\ref{Hunter, J. H., Jr. \& Fleck, R. C. 1982. Star formation: the
influence of velocity fields and turbulence. \apj 256: 505--513}

\ref{Hunter, J. H., Jr., Sandford, M. T., II, Whitaker, R. W., \&
Klein, R. I. 1986. Star formation in colliding gas flows. \apj 305:309--332}

\ref{Issa, M., MacLaren, I., \& Wolfendale, W. 1990. The
size-line width relation and the mass of molecular Hydrogen. \apj,
352, 132--138}

\ref{Kadomtsev, B.B. \& Petviashvili, V.I., 1973. Acoustic
Turbulence. Sov.\ Phys.\ Dokl.\ 18:115--116}

\ref{Kegel, W. H. 1989. The interpretation of correlation between
observed parameters of molecular clouds. \aa 225:517--520}

\ref{Kida, S. \& Orszag, S. A. 1990. Energy and spectral dynamics in
forced compressible turbulence. {\refit J. Sci. Computing} 5:85--125}

\ref{Klein, R. I. \& Woods, D. T. 1998. Bending Mode Instabilities and
Fragmentation in Interstellar Cloud Collisions: A Mechanism for
Complex Structure. \apj 497:777--799}

\ref{Kolmogorov, A. N. 1941. The local structure of turbulence in an
incompressible viscous fluid for very large Reynolds numbers. {\refit
Dokl. Akad. Nauk} 30:301-305}

\ref{Kornreich, P. \& Scalo, J. 1998. The Galactic shock pump. A
source of supersonic internal motions in the cool interstellar medium.}

\ref{Kraichnan, R.H.\ 1968. Lagrangian-history statistical theory for
Burgers' equation. {\refit Phys.\ Fluids} 11:265--277}

\ref{Kulsrud, R.M. and Pearce, W.P. 1969. The effect of wave-particle 
interactions on the propagation of cosmic rays. \apj 163:567--576}

\ref{Lada, E. A., Evans, N. J., II \& Falgarone, E. 1997. Physical
Properties of Molecular Cloud Cores in L1630 and Implications for Star
Formation. \apj 488:286--306}

\ref{Landau, L. D. \& Lifshitz, E. M. 1987. {\refit Fluid
Mechanics}. 2nd ed.\ (Oxford: Pergamon Press)}

\ref{Larson, R. B. 1981. Turbulence and star formation in molecular
clouds. \mnras 194:809--826}

\ref{Larson, R. B. 1995. Star formation in groups. \mnras 272:213--220}

\ref{Lazarian, A. 1995. Study of turbulence in HI using
radiointerferometers. \aa 293:507--520}

\ref{L\'eorat, J., Passot, T., \& pouquet, A.\ 1990. Influence of
supersonic turbulence on self-gravitating flows. \mnras 243:293--311}

\ref{Loren, R. B. 1989. The cobwebs of Ophiuchus. I. Strands of
(C-13)O - The mass distribution. \apj 338:902--924.}

\ref{Lou, Y.-Q. 1997. Gravitational collapse in the presence of a
finite-amplitude circularly polarized Alfv\'en wave. \mnras 279:L67--L71}

\ref{Mac Low, M.-M. 1998. The energy dissipation rate of supersonic,
magnetohydrodynamic turbulence in molecular clouds. \apj, submitted}

\ref{Mac Low, M.-M., Klessen, R. S. Burkert, \& Smith,
M. D. 1998. Kinetic energy decay of supersonic and super-Alfv\'enic
turbulence in star-forming clouds. {\refit Phys. Rev. Lett.} 80:2754--2757}

\ref{Maloney, P. 1988. The turbulent interstellar medium and
pressure-bounded molecular clouds. \apj 334:761--770}

\ref{McKee, C.~F. 1989.
        Photoionization-regulated star formation and the structure of
        molecular clouds. \apj 345:782--801}

\ref{McKee, C. F. \& Zweibel, E. G. 1992. On the virial theorem for
turbulent molecular clouds. \apj 399:551--562}

\ref{McKee, C.~F., Zweibel, E.~G., Goodman, A.~A., \& Heiles, C.
     1993. Magnetic fields in star-forming regions. In 
     {\refit Protostars and Planets III}, eds. E. H.  Levy \& J. I.
     Lunine (Tucson: University of Arizona Press), pp. 327--366}

\ref{McKee, C. F. \& Zweibel, E. G. 1995. Alfven Waves in Interstellar
Gasdynamics. \apj 440:686--696}

\ref{Miesch, M.S., and Bally, J. 1994.  Statistical analysis of
   turbulence in molecular clouds. \apj 429:645--671.

\ref{Mouschovias, T., and Spitzer, L.  Note on the collapse of 
magnetic interstellar clouds. \apj 210:326-327

\ref{Mouschovias, T. 1987. Star formation in magnetic interstellar
clouds. I. Interplay between theory and observation. In {\refit
Physical processes in interstellar clouds,} eds. G. E. Morfill \&
M. Scholer (Dordrecht: Reidel), p. 453--489}

\ref{Myers, P. C. 1978. A compilation of interstellar gas
properties. \apj 225:380--389}

\ref{Myers, P. C. \& Goodman, A. A. 1988a. Evidence for magnetic and
virial equilibrium in molecular clouds. \apjl 326:27--30}

\ref{Myers, P. C. \& Goodman, A. A. 1988b. Magnetic molecular clouds:
indirect evidence for magnetic support and ambipolar diffusion. \apj
329:392--405}

\ref{Myers, P.C., and Khersonsky, V.K. 1995. On magnetic turbulence in
 interstellar clouds.  \apj 442:186--196.}

\ref{Myers, P.C. 1995. Star forming molecular clouds. In {\refit Molecular 
Clouds and Star Formation}, eds. C. Yuan and Junhan You (Singapore: World 
Scientific), pp.\ 47--96}

\ref{Nordlund,\AA. \& Padoan, P. 1998. The density PDFs of supersonic
random flows. In {\refit Interstellar Turbulence, Proceedings of the
2nd Guillermo Haro Conference}, Eds. Franco, J. \& Carraminana, A.
(Cambridge:Cambridge University Press), in press }

\ref{Norman, C., \& Silk, J. 1980  Clumpy molecular clouds - A
  dynamic model self-consistently regulated by T Tauri star formation.
  \apj 238:158--174}

\ref{Norman, C. \& Ferrara, A. 1996. The Turbulent Interstellar
Medium: Generalizing to a Scale-dependent Phase Continuum. \apj 467:280--291}

\ref{Obukhov, A. M. 1941.  Of the distribution of energy in the
spectrum of turbulent flow. {\refit C. R. Acad. Sci. URSS} 32:19--}

\ref{\"Ogelman, H. B. \& Maran, S. P. 1976. The origin of OB
associations and extended regions of high-energy activity in the
Galaxy through supernova cascade processes. \apj 209:124--129}

\ref{Ostriker, E.C.  1997. Turbulence and magnetic fields in star formation.
In {\refit Star Formation, Near and Far}, Eds. S.Holt and L. Mundy.
(Woodbury NY: AIP Press), pp 51--63}

\ref{Ostriker, E.C., Gammie, C.F., and Stone, J.M. 1998.  Kinetic and
  structural evolution of self-gravitating, magnetized clouds:
  2.5-dimensional simulations of decaying turbulence.  \apj,
  in press}

\ref{Padoan, P., Jones, B.J.T., Nordlund, \AA. 1997. Supersonic
   Turbulence in the Interstellar Medium: Stellar Extinction
   Determinations as Probes of the Structure and Dynamics of Dark
   Clouds \apj 474:730--734}

\ref{Padoan, P., Nordlund, \AA. \& Jones, B. J. T. 1997. The
universality of the stellar initial mass function. \mnras 288:145--152}

\ref{Padoan, P., \& Nordlund, \AA. 1998a. A Super-Alfv\'enic Model for
Dark Clouds. \apj, submitted.}

\ref{Padoan, P., \& Nordlund, \AA. 1998b. Super-Alfv\'enic turbulent
fragmentation in molecular clouds. In {\refit Interstellar Turbulence,
Proceedings of the 2nd Guillermo Haro Conference}, Eds. Franco, J. \&
Carraminana, A. (Cambridge:Cambridge University Press), in press}

\ref{Padoan, P., Juvela, M., Bally, J., \& Nordlund,
\AA. 1998. Synthetic molecular clouds from supersonic MHD and non-LTE
radiative transfer calculations. \apj 504:300--313}

\ref{Passot, T., Pouquet, A. \& Woodward, P. 1988. The plausibility of
Kolmogorov-type spectra in molecular clouds. \aa 197: 228--234}

\ref{Passot, T., \VS, E., \& Pouquet, A. 1995. A turbulent model for
the interstellar medium. II. Magnetic fields and rotation. \apj 455:536--555}

\ref{Passot, T. \& \VS, E. 1998. Density probability distribution in
one-dimensional gas dynamics. {\refit Phys. Rev. E}, 58:4501--4510}

\ref{Peng, R., Langer, W. D., Velusamy, T., Kuiper, T. B. H., \&
Levin, S. 1998. Low-mass clumps in TMC-1: Scaling laws in the
small-scale regime. \apj 197:842--849}

\ref{Peters, W. L. \& Bash, F. N. 1987. The correlation of spiral arm
molecular clouds with atomic hydrogen. \apj 317:646--652}

\ref{Plume, R., Jaffe, D. T., Evans, N. J., II, Mart\'in-Pintado, J.,
\& G\'omez-Gonz\'alez, J. 1997.Dense Gas and Star Formation:
Characteristics of Cloud Cores Associated with Water Masers. \apj 476:730--749}

\ref{Pouquet, A., Passot, T. \& L\'eorat, J. 1990. Numerical Simulations of
Compressible Flows. Proceedings of the IAU symposium 147. In {\refit
Fragmentation of Molecular Clouds and Star Formation},
eds. E. Falgarone, F. Boulanger and G. Duvert. (Dordrecht:Kluwer), p. 101--118}

\ref{Pratap, P., Dickens, J. E., Snell, R. L., Miralles, M. P.,
Bergin, E. A., Irvine, W. M., \& Schloerb, F. P. 1997. A Study of the
Physics and Chemistry of TMC-1. \apj 486:862--885}

\ref{Rosen, A., Bregman, J. N., \& Norman, M. L. 1993. Hydrodynamical
simulations of star-gas interactions in the interstellar medium with
an external gravitational potential. \apj 413:137--149}

\ref{Scalo, J 1985. Fragmentation and hierarchical structure in the
interstellar medium. In {\refit Protostars \& Planets II}, eds. Black,
D. C. \& Matthews, M. S. (Tucson: University of Arizona Press), p. 201--296}

\ref{Scalo, J. 1987. Theoretical approaches to interstellar
turbulence. In {\refit Interstellar Processes\/},
eds. D. J. Hollenbax]ch and H. A. Thronson, Jr. (Dordrecht:Reidel), p.
349--392}

\ref{Scalo, J. 1990. Perception of interstellar structure. Facing
complexity. In{\refit Physical processes in fragmentation and star
formation}, eds. R. Capuzzo-Dolcetta, C. Chiosi and A. di Fazio
(Dordrecht: Kluwer), p. 151--177}

\ref{Scalo, J, \VS, E., Chappell, D., \& Passot, T. 1998. On the
Probability Density Function of Galactic Gas. I. Numerical Simulations
and the Significance of the Polytropic Index. \apj 504:835--853}

\ref{Shaya, E. J. \& Federman, S. R. 1987. The HI distribution in
clouds within galaxies. \apj 319:76--83}

\ref{Shu, F. H., Adams, F. C., \& Lizano, S. 1987. Star formation in
molecular clouds.: Observation and theory. {\refit
{Ann. Rev. Astron. Ap.} 25:23--81}

\ref{Simon, M. 1997. Clustering of young stars in Taurus, Ophiucus,
and the Orion Trapezium. \apjl 482:81--84}

\ref{Slavin, J. D. \& Cox, D. P. 1992. Completing the evolution of
supernova remnants and their bubbles. \apj 392:131--144}

\ref{Solomon, P. M  \& Sanders, D. B. 1985. Star formation in a
Galactic context: The location and properties of molecular clouds. In
{\refit Protostars \& Planets II}, eds. Black, D. C. \& Matthews,
M. S. (Tucson: University of Arizona Press), p. 59--80}

\ref{Spitzer, L., Jr. \& Savedoff, M. P. 1950. The temperature of
interstellar matter. III. \apj 111:593--608}

\ref{Stevens, I. R., Blondin, J. M \& Pollock, A. M. T. 1992. 
Colliding winds from early-type stars in binary systems.\apj 386:265--287} 

\ref{Stone, J.M. 1998.  Direct Numerical Simulations of Compressible
  MHD Turbulence.  In {\refit Interstellar Turbulence, Proceedings of
  the 2nd Guillermo Haro Conference}, Eds. Franco, J. \& Carraminana,
  A.  (Cambridge:Cambridge University Press)}

\ref{Stone, J.M., Ostriker, E.C., and Gammie, C.F. 1998.  Dissipation in
compressible MHD turbulence.  \apjl, in press}

\ref{Tohline, J. E, Bodenheimer, P. H., \& Christodoulou,
D. M. 1987. The crucial role of cooling in the making of molecular
clouds and stars. \apj 322:787--794}

\ref{Tomisaka, K., Ikeuchi, S., and Nakamura T. 1988. Equilibria and
    evolutions of magnetized, rotating, isothermal clouds. II - The
    extreme case: Nonrotating clouds.  \apj 335:239--262}

\ref{Torrelles, J. M., Rodr\'iguez, L. F., Cant\'o, J., Marcaide, J.,
\& Gyulbudaghian, A. L. 1983. A search for molecular outflows
associated with peculiar nebulosities and regions of star
formation. {\refit Rev. Mex. Astron. Astrof.} 8:147--154}

\ref{Troland, T. H., Crutcher, R. M., Goodman, A. A., Heiles, C.,
  Kazes, I., \& Myers, P. C. 1996. The Magnetic Fields in the
  Ophiuchus and Taurus Molecular Clouds.\apj 471:302--307

\ref{V\'azquez, E. C. \& Scalo, J. 1989. Evolution of the star
formation rate in galaxies with increasing densities. \apj 343:644--658} 

\ref{\VS, E. 1994. Hierarchical structure in nearly pressureless
flows as a consequence of self-similar statistics. \apj 423:681--692}

\ref{\VS, E. \& Gazol, A. 1995. Gravitational instability in turbulent,
non-uniform media. \aa 303:204--210}

\ref{\VS, E., Passot, T.,\& Pouquet, A. 1995. A Turbulent Model for the
interstellar medium. I. Threshold star formation and
self-gravity. \apj 441:702--725}

\ref{\VS, E., Passot, T.,\& Pouquet, A. 1996.Influence of
cooling-induced compressibility on the structure of turbulent flows
and gravitational collapse. \apj 473:881--893} 

\ref{\VS, E., Ballesteros-Paredes, J. \& Rodr\'iguez, L. F. 1997. A
search for Larson-type relations in numerical simulations of the ISM. Evidence
for non-constant column densities. \apj 474:292--307}

\ref{\VS, E., Cant\'o, J. y Lizano, S. 1998. Does turbulent pressure
behave as a logatrope? \apj 492:596--602} 

\ref{\VS, E. \& Passot, T. 1998. Turbulence as an organizing agent in
the ISM. In {\refit Interstellar Turbulence, Proceedings of the 2nd Guillermo
Haro Conference}, Eds. Franco, J. \& Carraminana,
A. (Cambridge:Cambridge University Press), in press} 

\ref{\VS, E. 1998. Turbulence in Molecular Clouds. In {\refit
Chemistry, Physics, and Observations of Molecules in Space. Proc. of
the 1996 INAOE Summer School of Millimeter-Wave Astronomy},
eds. W. F. Wall, A. Carraminana, L. Carrasco, and P. F. Goldsmith, in press} 

\ref{Vishniac, E. T. 1983. The dynamic and gravitational instabilities
of spherical shocks. \apj 274:152--167}

\ref{Vishniac, E. T. 1994. Nonlinear instabilities in shock-bounded
slabs. \apj 428:186--208}

\ref{Wannier, P. G., Lichten, S. M., Andersson, B.-G., \& Morris,
M. 1991. Warm neutral halos around molecular clouds. II. H I and CO (J
= 1-0) observations. \apjs 75:987--998}

\ref{Whitworth, A. 1996. The structure of the neutral interstellar
medium: A theory of interstellar turbulence. In {\refit Unsolved
problems of the Milky Way}, eds. , p. 591--596}

\ref{Williams, J. P., Blitz, L., \& Stark, A. A. 1995. The density
structure in the Rosette molecular cloud: Signposts of evolution. \apj
451: 252--274}

\ref{Wilson, T. L. \& Walmsley, C. M. 1989. Small-scale clumping in molecular
clouds. {\refit Astron. Astrphys. Rev.} 1:141-176.}

\ref{Wolfire, M. G., Hollenbach, D., McKee, C. F., Tielens,
A. G. G. M., \& Bakes, E. L. O. 1995. The neutral atomic phases of the 
interstellar medium. \apj, 443:152--168}





\vfill\eject
\null

\vskip .5in
\centerline{\bf FIGURE CAPTIONS}
\vskip .25in
  
\caption{Figure 1.\capskip Sample density, velocity, and magnetic field 
structure in two turbulent cloud simulations.  Density contours 
are logarithmic, with dark blue running from the mean density to ten times
the mean, and light blue 10-100 times the mean;  magnetic field lines are
shown in green, and the velocity field is shown in red.  For both models,
the Mach number is ten;  for the upper (lower) figure, $\beta=0.01$ (0.1),
corresponding to stronger (weaker) mean magnetic fields.  See Ostriker et al.\ 
(1998) for more details.
}

\vfil\eject
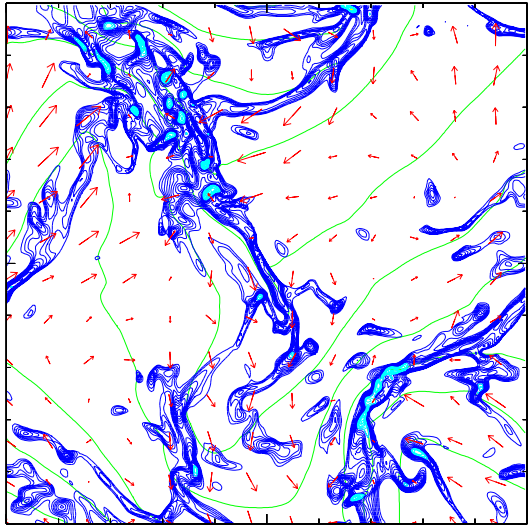
\end